\documentclass[10pt]{article}
\usepackage{graphicx}

\usepackage{latexsym}
\usepackage{amssymb}
\usepackage{amsfonts}
\usepackage{amsmath}
\usepackage{theorem}
\newtheorem{theorem}{Theorem}

\newtheorem{definition}{Definition}

\newtheorem{proposition}{Proposition}
\makeatletter
\newcommand{\contraction}[5][1ex]{%
  \mathchoice
    {\contraction@\displaystyle{#2}{#3}{#4}{#5}{#1}}%
    {\contraction@\textstyle{#2}{#3}{#4}{#5}{#1}}%
    {\contraction@\scriptstyle{#2}{#3}{#4}{#5}{#1}}%
    {\contraction@\scriptscriptstyle{#2}{#3}{#4}{#5}{#1}}}%
\newcommand{\contraction@}[6]{%
  \setbox0=\hbox{$#1#2$}%
  \setbox2=\hbox{$#1#3$}%
  \setbox4=\hbox{$#1#4$}%
  \setbox6=\hbox{$#1#5$}%
  \dimen0=\wd2%
  \advance\dimen0 by \wd6%
  \divide\dimen0 by 2%
  \advance\dimen0 by \wd4%
  \vbox{%
    \hbox to 0pt{%
      \kern \wd0%
      \kern 0.5\wd2%
      \contraction@@{\dimen0}{#6}%
      \hss}%
    \vskip 0.5ex
    \vskip\ht2}}

\newcommand{\contraction@@}[3][0.05em]{%
  \hbox{%
    \vrule width #1 height 0pt depth #3%
    \vrule width #2 height 0pt depth #1%
    \vrule width #1 height 0pt depth #3%
    \relax}}
\makeatother

\begin{document}

\title{\bf Criticality in diluted ferromagnet}
\author{Elena Agliari\footnote{Theoretische Polymerphysik, Universit\"{a}t
Freiburg, Germany
 {\tt<elena.agliari@physik.uni-freiburg.de>}},\ Adriano Barra\footnote{Dipartimento di Fisica, Universit\`a di Roma ``La
Sapienza'' and Dipartimento di Matematica, Universit\`a di
Bologna, Italy, {\tt<Adriano.Barra@roma1.infn.it>}},\ Federico
Camboni\footnote{Dipartimento di Fisica, Universit\`a di Roma ``La
Sapienza'', Italy, {\tt<Federico.Camboni@uniroma1.it>}}}

\def\be{\begin{equation}}
\def\ee{\end{equation}}
\def\bc{\begin{center}}
\def\ec{\end{center}}

\maketitle

\begin{abstract}
We perform a detailed study of the critical behavior of the mean
field diluted Ising ferromagnet by analytical and numerical tools.
We obtain self-averaging for the magnetization and write down an
expansion for the free energy close to the critical line. The
scaling of the magnetization is also rigorously obtained and
compared with extensive Monte Carlo simulations. We explain the
transition from an ergodic region to a non trivial phase by
commutativity breaking of the infinite volume limit and a suitable
vanishing field. We find full agreement among theory, simulations
and previous results.
\end{abstract}

\section{Introduction}

In the past few years a match among the study of systems defined
on lattices by means of statistical mechanics \cite{leshouches}
and the study of networks  by means of graph theory \cite{guido}
gave origin to very interesting models as the small world magnets
\cite{SW}\cite{scazzo1} or the scale free networks \cite{guido2}.
However a complete analysis starting from the simple fully
connected mean field Ising model \cite{abarra}\cite{ellis} up to
these recent complex models \cite{barabasi} is still not complete
(even thought several important steps have been obtained, examples
being \cite{almeida}\cite{bovier}\cite{gallo}) and important
models have not yet been taken into account. Among these a certain
role is played by the mean field diluted Ising model: an
Erdos-Renyi networks \cite{erdos} which has spins as nodes and
their interactions as links. The interactions are encoded in a
matrix connecting pairs of spins which, when the connectivity
allows the link to be present, shares the same value for all the
couples.
\newline
Despite their easy formalization \cite{gulielmo}, diluted
ferromagnets are poorly investigated by rigorous tools
\cite{montanari}. Inspired by a recent work on these systems in
which the authors presented a detailed analysis of the ergodic
region and the zero temperature line \cite{dsg} we extend recent
techniques developed in a series of papers
\cite{ac}\cite{barra1}\cite{barra3}\cite{barra5} to this model
with the aim of analyzing its critical behavior. We systematically
develop the interpolating cavity field method \cite{guerra4} and
use it to sketch the derivation of a free energy expansion: the
higher the order of the expansion, the deeper we could go beyond
the ergodic region. Within this framework we perform a detailed
analysis of the scaling of magnetization (and susceptibility) at
the critical line. The critical exponents turn out to be the
classical ones. At the end we perform extensive Monte Carlo (MC)
simulations for different graph sizes and bond concentrations and
we compare results with theory. Indeed, also numerically, we
provide evidence that the universality class of the diluted Ising
model is independent of dilution. In fact the critical exponents
we measured are consistent with those pertaining to the
Curie-Weiss model, in agreement with analytical results. The
critical line is also well reproduced.
\newline
The paper is organized as follows: In Section (\ref{laA}) we
describe the model, in Section (\ref{cavatappi}) we introduce the
cavity field technique, which constitutes the framework we are
going to use in Section (\ref{laC}) to investigate the free energy
of the system at general values of temperature and dilution.
Section (\ref{critico}) deals with the criticality of the model;
there we find the critical line and the critical behavior of the
main order parameter, i.e. magnetization, we provide its
self-averaging and work out a picture by which we explain the
breaking of the ergodicity. Section (\ref{numerics}) is devoted to
numerical investigations, especially focused on criticality.
Finally, Section (\ref{conclusions}) is left for outlook and
conclusions.

\section{Model and notations}\label{laA}

Given $N$ points and families $\{i_{\nu},j_{\nu}\}$ of i.i.d
random variables uniformly distributed on these points, the
(random) Hamiltonian of the diluted Curie-Weiss model is defined
on Ising $N$-spin configurations
$\sigma=(\sigma_1,\ldots,\sigma_N)$ through
\begin{equation}\label{ham}
H_{N}(\sigma,\alpha)=-\sum_{\nu=1}^{P_{\alpha N}}
\sigma_{i_{\nu}}\sigma_{j_{\nu}}\ ,
\end{equation}
where $P_{\zeta}$ is a Poisson random variable with mean $\zeta$
 and $\alpha>1/2$ is the connectivity. The expectation with respect
to all the (\emph{quenched}) random variables defined so far will
be denoted by $\mathbb{E}$, while the Gibbs expectation at inverse
temperature $\beta$ with respect to this Hamiltonian will be
denoted by $\Omega$, and depends clearly on $\alpha$ and $\beta$.
We also define $\langle \cdot \rangle = \mathbb{E}\Omega(\cdot)$.
The pressure, i.e. minus $\beta$ times the free energy, is by
definition
$$
A_{N}(\alpha)= \frac{1}{N}\mathbb{E}\ln Z_N(\beta) =
\frac1N\mathbb{E}\ln\sum_{\sigma}\exp(-\beta
H_{N}(\sigma,\alpha))\
$$
where we implicitly introduced the partition function $Z_N(\beta)$
too. When we omit the dependence on $N$ we mean to have taken the
thermodynamic limit which we assume to exist for all the
observables we deal with, in particular for the free energy
\cite{montanari}\cite{bianchi}\cite{GT}
 (however we will look for a more firm  ground on this
point by numerical investigation in sec.(\ref{numerics})). The
quantities encoding the thermodynamic properties of the model are
 the overlaps, which are defined on several configurations
(\emph{replicas}) $\sigma^{(1)},\ldots,\sigma^{(n)}$ by
$$
q_{1\cdots
n}=\frac1N\sum_{i=1}^{N}\sigma^{(1)}_{i}\cdots\sigma^{(n)}_{i} .
$$
Particular attention must be payed on $ q_1 = m = N^{-1}\sum_{i}^N
\sigma_i$ which is called {\em magnetization}.
\newline
When dealing with several replicas, the Gibbs measure is simply
the product measure, with the same realization of the quenched
variables, but the expectation $\mathbb{E}$ destroys the
factorization.
\newline
Sometimes for the sake of simplicity we will call $\theta =
\tanh(\beta)$.

\section{Interpolating with the cavity field}\label{cavatappi}

In this section at first we introduce the cavity field technique
on the line of \cite{barra1} by expressing the Hamiltonian of a
system made of $N+1$ spins through the Hamiltonian of N spins by
scaling the connectivity degree $\alpha$ and neglecting vanishing
terms in N as follows \be H_{N+1}(\alpha) =
-\sum_{\nu=1}^{P_{\alpha (N+1)}} \sigma_{i_\nu}\sigma_{j_\nu}
\quad \sim \quad -\sum_{\nu=1}^{P_{\tilde{\alpha} N}}
\sigma_{i_\nu}\sigma_{j_\nu} - \sum_{\nu=1}^{P_{2\tilde{\alpha }}}
\sigma_{i_\nu}\sigma_{N+1} \ee such that we can use the more
compact expression
\begin{equation}\label{hsplit}
 H_{N+1}(\alpha) \sim H_{N}(\tilde{\alpha}) +
\hat{H}_{N}(\tilde{\alpha})\sigma_{N+1}
\end{equation}
with \be\label{decompo} \tilde{\alpha} = \frac{N}{N+1}\alpha
\stackrel{N\rightarrow \infty}{\longrightarrow} \alpha, \qquad
\hat{H}_{N}(\tilde{\alpha}) = - \sum_{\nu=1}^{P_{2\tilde{\alpha}}}
\sigma_{i_\nu}. \ee
\medskip
So we see that we can express the Hamiltonian of $N+1$ particles
via the one of $N$ particles, paying two prices: the first is a
rescaling in the connectivity (vanishing in the thermodynamic
limit), the second being an added term, which will be encoded, at
the level of the thermodynamics, by a suitably cavity function as
follows: let us introduce an interpolating parameter $t \in [0,1]$
 and the cavity function $\Psi_N(\tilde{\alpha},t)$ given by
\begin{eqnarray}
\Psi(\tilde{\alpha},\beta;t) &=& \lim_{N\rightarrow \infty
}\Psi_N(\tilde{\alpha}, \beta; t)  \\
\nonumber &&\lim_{N\rightarrow \infty } \mathbb{E}\Big[\ln
\frac{\sum_{\{\sigma\}} e^{\beta\sum_{\nu=1}^{P_{\tilde{\alpha}
N}} \sigma_{i_\nu}\sigma_{j_\nu} + \beta
\sum_{\nu=1}^{P_{2\tilde{\alpha }t}} \sigma_{i_\nu}}}
{\sum_{\sigma} e^{\beta\sum_{\nu=1}^{P_{\tilde{\alpha} N}}
\sigma_{i_\nu}\sigma_{j_\nu}}}\Big] = \lim_{N\rightarrow \infty }
\mathbb{E}\Big[\ln \frac{Z_{N,t}(\tilde{\alpha},\beta)}
{Z_{N}(\tilde{\alpha},\beta)}\Big].
\end{eqnarray}
The three terms appearing in the decomposition (\ref{hsplit}) give
rise to the structure of the following theorem:
\begin{theorem}\label{main}
In the $N\rightarrow \infty$ limit,  the free energy per spin is
allowed to assume the following representation \be A(\alpha,\beta)
= \ln2 - \alpha \frac{\partial A(\alpha,\beta)}{\partial\alpha} +
\Psi(\alpha,\beta;t=1) \ee
\end{theorem}
\medskip
\textbf{Proof}
\newline
Consider the $N+1$ spin partition function $Z_{N+1}(\alpha,\beta)$
and let us split it as suggested by eq. ($\ref{hsplit})$
\begin{eqnarray}\label{zecca}
Z_{N+1}(\alpha,\beta) &=& \sum_{\{\sigma_{N+1}\}} e^{-\beta
H_{N+1}(\alpha)} \sim \sum_{\{\sigma_{N+1}\}} e^{-\beta
H_{N}(\tilde{\alpha}) -
\beta\hat{H}_{N}(\tilde{\alpha})\sigma_{N+1}}
\\ \nonumber
 &=& \sum_{\{\sigma_{N+1}\}}
e^{\beta\sum_{\nu=1}^{P_{\tilde{\alpha} N}}
\sigma_{i_\nu}\sigma_{j_\nu} + \beta
\sum_{\nu=1}^{P_{2\tilde{\alpha }}} \sigma_{i_\nu}\sigma_{N+1}} =
2 \sum_{\{\sigma_{N}\}} e^{\beta\sum_{\nu=1}^{P_{\tilde{\alpha}
N}} \sigma_{i_\nu}\sigma_{j_\nu} + \beta
\sum_{\nu=1}^{P_{2\tilde{\alpha }}} \sigma_{i_\nu}}
\end{eqnarray}
\medskip
where the factor two appears because of the sum over the hidden
$\sigma_{N+1}$ variable. Defining a perturbed Boltzmann state
$\tilde{\omega}$ (and  its replica product
$\tilde{\Omega}=\tilde{\omega}\times...\times\tilde{\omega}$) as
$$
\tilde{\omega}(g(\sigma)) =
\frac{\sum_{\{\sigma_{N}\}}g(\sigma)e^{-\beta
H_N(\tilde{\alpha})}} {\sum_{\{\sigma_{N}\}}e^{-\beta
H_N(\tilde{\alpha})}}, \ \ \ \ \ \tilde{\Omega}(g(\sigma)) =
\prod_i\tilde{\omega}^{(i)}(g(\sigma^{(i)}))
$$
where the tilde takes into account the shift in the connectivity
$\alpha \rightarrow \tilde{\alpha}$ and multiplying and dividing
the r.h.s. of eq.(\ref{zecca}) by $Z_N(\tilde{\alpha},\beta)$, we
obtain
\begin{equation}\label{dinox}
Z_{N+1}(\alpha,\beta) = 2 Z_N(\tilde{\alpha},\beta)
\tilde{\omega}(e^{\beta\sum_{\nu=1}^{P_{2\tilde{\alpha }}}}).
\end{equation}
\medskip
Taking now the logarithm of both sides of eq.(\ref{dinox}),
applying the average $\mathbb{E}$ and subtracting the quantity
$\mathbb[\ln Z_{N+1}(\tilde{\alpha},\beta)]$, we get
\begin{equation}\label{latte}
\mathbb{E}[\ln Z_{N+1}(\alpha,\beta)] - \mathbb{E}[\ln
Z_{N+1}(\tilde{\alpha},\beta)] = \ln2 + \mathbb{E}\Big[\ln
\frac{Z_N(\tilde{\alpha},\beta)}{Z_{N+1}(\tilde{\alpha},\beta)}\Big]
+ \Psi_N(\tilde{\alpha},\beta;t=1)
\end{equation}
\medskip
in the large $N$ limit the l.h.s. of eq.(\ref{latte}) becomes
\begin{eqnarray}
\mathbb{E}[\ln Z_{N+1}(\alpha,\beta)] - \mathbb{E}[\ln
Z_{N+1}(\tilde{\alpha},\beta)] &=& \\ \nonumber (\alpha -
\tilde{\alpha}) \frac{\partial}{\partial\alpha}\mathbb{E}[\ln
Z_{N+1}(\alpha,\beta)]
 &=& \alpha \frac{1}{N+1}\frac{\partial}{\partial\alpha}
\mathbb[\ln Z_{N+1}(\alpha,\beta)] = \alpha \frac{\partial
A_{N+1}(\alpha,\beta)} {\partial\alpha}
\end{eqnarray}
\medskip
then by considering the thermodynamic limit the thesis follows.
(Actually we still do not have a complete proof of the existence
of the thermodynamic limit but we will provide strong numerical
evidences in Section \ref{numerics}) $\Box$

\bigskip

Hence, we can express the free energy via the energy and the
cavity function. While it is well known how to deal with the
energy \cite{dsg}, the same can not be stated for the cavity
function, and we want to develop its expansion via suitably chosen
overlap monomials in a spirit close to the stochastic stability
\cite{ac}\cite{cg2}\cite{parisiSS}, such that, at the end, we will
not have the analytical solution for  the free energy in the whole
$(\alpha,\beta)$ plane, but we will manage its expansion close
(immediately below) to the critical line. To see how the machinery
works, let us start by giving some definitions and proving some
simple theorems:
\begin{definition}
We define the t-dependent Boltzmann state $\tilde{\omega}_t$ as
\begin{equation}\label{dante}
\tilde{\omega}_t(g(\sigma)) = \frac{1}{Z_{N,t}(\alpha,\beta)}
\sum_{\{\sigma\}}g(\sigma) e^{\beta\sum_{\nu=1}^{P_{\tilde{\alpha}
N}} \sigma_{i_\nu}\sigma_{j_\nu} + \beta
\sum_{\nu=1}^{P_{2\tilde{\alpha }t}} \sigma_{i_\nu}}
\end{equation}
where $Z_{N,}(\alpha\beta)$ extends the classical partition
function in the same spirit of the numerator of eq.(\ref{dante}).
\end{definition}
As we will often deal with several overlap monomials let us divide
them among two big categories:
\begin{definition} We can split the class of monomials of the
order parameters in two families:
\begin{itemize}

\item We define {\itshape filled} or equivalently {\itshape stochastically
stable} all the overlap monomials built by an even number of the
same replicas (i.e. $q_{12}^2$,\ $m^2$,\ $q_{12}q_{34}q_{1234}$).

\item We define {\itshape fillable} or equivalently {\itshape saturable}
all the overlap monomials which are not stochastically stable
(i.e. $q_{12}$,\ $m$,\ $q_{12}q_{34}$)
\end{itemize}
\end{definition}
We are going to show three theorems that will play a guiding role
for our expansion: as this approach has been deeply developed in
similar contexts (as fully connected Ising model \cite{abarra} or
fully connected spin glasses \cite{barra1} or diluted spin glasses
\cite{barra4}, which are the boundaries of the model of this
paper) we will not show all the details of the proof, but we
sketch them as they are really intuitive. The interested reader
can deepen this point by looking at the original works.
\begin{theorem}\label{ciccia}
For large $N$, setting $t=1$ we have \be
\tilde{\omega}_{N,t}(\sigma_{i_1}\sigma_{i_2}...\sigma_{i_n}) =
\tilde{\omega}_{N+1}(\sigma_{i_1}\sigma_{i_2}...\sigma_{i_n}\sigma_{N+1}^n)
+ O(\frac{1}{N})\ee such that in the thermodynamic limit, if
$t=1$, the Boltzmann average of a fillable multi-overlap monomial
turns out to be the Boltzmann average of the corresponding filled
multi-overlap monomial.
\end{theorem}
\begin{theorem}\label{saturabili}
Let $Q_{2n}$ be a fillable monomial of the overlaps (this means
that there exists a multi-overlap $q_{2n}$ such that
$q_{2n}Q_{2n}$ is filled). We have \be
\lim_{N\rightarrow\infty}\lim_{t\rightarrow1} \langle Q_{2n}
\rangle_t = \langle q_{2n}Q_{2n} \rangle \ee \bc (examples: for $N
\rightarrow \infty$ we get $\langle m_1 \rangle_t \rightarrow
\langle m_1^2 \rangle,\ \langle q_{12} \rangle_t \rightarrow
\langle q_{12}^2 \rangle,\ \langle q_{12}q_{34} \rangle_t
\rightarrow \langle q_{12}q_{34}q_{1234} \rangle$) \ec
\end{theorem}
\begin{theorem}\label{saturi}
In the $N\rightarrow\infty$ limit the averages
$\langle\cdot\rangle$ of the filled monomials are t-independent in
$\beta$ average.
\end{theorem}
\medskip
\textbf{Proof}
\newline
In this sketch we are going to show how to get Theorem
(\ref{ciccia}) in some details; It automatically has as a
corollary Theorem (\ref{saturabili}) which ultimately gives, as a
simple consequence when applied on filled monomials, Theorem
(\ref{saturi}).
\newline
Let us assume for a generic overlap correlation function $Q$, of
$s$ replicas, the following representation
$$
Q = \prod_{a=1}^s\sum_{i_l^a}\prod_{l=1}^{n^a}\sigma_{i_l^a}^a
I(\{i_l^a \})
$$
where $a$ labels the replicas, the internal product takes into
account the spins (labeled by $l$) which contribute to the
{\itshape a}-part of the overlap $q_{a,a'}$ and runs to the number
of time that the replica $a$ appears in $Q$, the external product
takes into account all the contributions of the internal one and
the $I$ factor fixes the constraints among different replicas in
$Q$; so, for example, $Q=q_{12}q_{23}$ can be decomposed in this
form noting that $s=3$, $n^1=n^3=1,n^2=2$,
$I=N^{-2}\delta_{i_1^1,i_1^3}\delta_{i_1^2,i_2^3}$, where the
$\delta$ functions fixes the links between replicas $1,2
\rightarrow q_{1,2}$ and $2,3 \rightarrow q_{2,3}$. The averaged
overlap correlation function is
$$
\langle Q \rangle_t = \mathbb{E}\sum_{i_l^a}I(\{i_l^a
\})\prod_{a=1}^s \omega_{t}(\prod_{l=1}^{n^a}\sigma_{i_l^a}^a).
$$
Now if $Q$ is a
fillable polynomial, and we evaluate it at $t=1$, let us decompose
it, using the factorization of the $\omega$ state on different
replica, as
$$ \langle Q \rangle_t = \mathbb{E}\sum_{i_l^a,i_l^b}I(\{i_l^a \}, \{i_l^b \})\prod_{a=1}^u
\omega_a ( \prod_{l=1}^{n^a}\sigma_{i_l^a}^a) \prod_{b=u}^s
\omega_b ( \prod_{l=1}^{n^b}\sigma_{i_l^b}^b),
$$
where $u$ stands
for the number of the unfilled replicas inside the expression of
$Q$. So we split the measure $\Omega$ into two different subset
$\omega_{a}$ and $\omega_{b}$: in this way the replica belonging
to the $b$ subset are always in even number, while the ones in the
$a$ subset are always odds. Applying the gauge $\sigma_i^a
\rightarrow \sigma_i^a\sigma_{N+1}^a, \forall i \in (1,N)$ the
even measure is unaffected by this transformation
$(\sigma_{N+1}^{2n} \equiv 1)$ while the odd measure takes a
$\sigma_{N+1}$ inside the Boltzmann measure.
$$
\langle Q \rangle = \sum_{i_l^a,i_l^b}I(\{i_l^a \}, \{i_l^b \}) \prod_{a=1}^u
\omega ( \sigma_{N+1}^a \prod_{l=1}^{n^a}\sigma_{i_l^a}^a)
\prod_{b=u}^s \omega (
\sigma_{N+1}^b\prod_{l=1}^{n^b}\sigma_{i_l^b}^b)
$$
At the end we can replace in the last expression the subindex
$N+1$ of $\sigma_{N+1}$ by $k$ for any $k \neq \{ i_l^a \}$ and
multiply by one as $1=N^{-1}\sum_{k=0}^N$. Up to orders $O(1/N)$,
which go to zero in the thermodynamic limit, we have the proof.
$\Box$
\medskip
\newline
It is now immediate to understand that the effect of Theorem
(\ref{ciccia}) on a fillable overlap monomial is to multiply it by
its missing part to be filled (Theorem \ref{saturabili}), while it
has no effect if the overlap monomial is already filled (Theorem
\ref{saturi}) because of the Ising spins (i.e. $\sigma_{N+1}^{2n}
\equiv 1 \ \forall n \in \mathbb{N}$).

\bigskip

Now the plan is as follows:  We calculate the $t$-streaming of the
$\Psi$ function in order to derive it and then integrate it back
 once we have been able to express it as an expansion in power series of $t$ with
stochastically stable overlaps as coefficients. At the end we free
the perturbed Boltzmann measure by setting $t=1$ and in the
thermodynamic limit we will have the expansion holding with the
correct statistical mechanics weight.
\medskip
\begin{eqnarray}\label{giura} \frac{\partial\Psi(\tilde{\alpha}, \beta, t)}{\partial t}
&=& \frac{\partial} {\partial t}\mathbb{E}[\ln \tilde{\omega}
(e^{\beta \sum_{\nu=1}^{P_{2\tilde{\alpha }t}} \sigma_{i_\nu}})]
\\ \nonumber
&=& 2\tilde{\alpha}\mathbb{E}[\ln \tilde{\omega} (e^{\beta
\sum_{\nu=1}^{P_{2\tilde{\alpha }t}} \sigma_{i_\nu} +
\beta\sigma_{i_0}})] - 2\tilde{\alpha}\mathbb{E}[\ln
\tilde{\omega} (e^{\beta \sum_{\nu=1}^{P_{2\tilde{\alpha }t}}
\sigma_{i_\nu}})] = 2\tilde{\alpha}\mathbb{E}[\ln \tilde{\omega}_t
(e^{\beta\sigma_{i_0}})]
\end{eqnarray}
\medskip
using now the equality $e^{\beta\sigma_{i_0}}=\cosh\beta +
\sigma_{i_0}\sinh\beta$, we can write the r.h.s. of
eq.(\ref{giura}) as
\medskip
$$
\frac{\partial\Psi(\tilde{\alpha}, \beta, t)}{\partial t} =
2\tilde{\alpha}\mathbb{E}[\ln \tilde{\omega}_t (\cosh\beta +
\sigma_{i_0}\sinh\beta )] = 2\tilde{\alpha}\log\cosh\beta -
2\tilde{\alpha}\mathbb{E}[\ln(1 +
\tilde{\omega}_t(\sigma_{i_0})\theta)].
$$
We can expand the function $\log(1 + \tilde{\omega}_t \theta)$ in
powers of $\theta$, obtaining
\begin{equation}\label{sedici}
\frac{\partial\Psi(\tilde{\alpha},t)}{\partial t} =
2\tilde{\alpha}\ln\cosh\beta -
2\tilde{\alpha}\sum_{n=1}^{\infty}\frac{(-1)^n}{n} \theta^n\langle
q_{1,...,n} \rangle_t.
\end{equation}
We learn by looking at eq.(\ref{sedici}) that the derivative of
the cavity function is built by non-stochastically stable overlap
monomials, and their averages do depend on $t$ making their
$t$-integration non trivial (we stress that all the fillable terms
are zero when evaluated at $t=0$ due to the gauge invariance of
the model). We can escape this constraint by iterating them again
and again (and then integrating them back too) because their
derivative, systematically, will develop stochastically stable
terms, which turn out to be independent by the interpolating
parameter and their integration is straightforwardly polynomial.
To this task we introduce the following
\begin{proposition}\label{stream}
Let $F_s$ be a function of s replicas. Then the following
streaming equation holds
\begin{eqnarray}
\frac{\partial\langle F_s \rangle_{t,\tilde{\alpha}}}{\partial t}
&=& 2\tilde{\alpha}\theta [\sum_{a=1}^s\langle F_s \sigma_{i_0}^a
\rangle_{t,\tilde{\alpha}} - s \langle F_s \sigma_{i_0}^{s+1}
\rangle_{t,\tilde{\alpha}}] \quad
\\ \nonumber
&+& \quad 2\tilde{\alpha}\theta^2 [ \sum_{a<b}^{1,s}\langle F_s
\sigma_{i_0}^a\sigma_{i_0}^b \rangle_{t,\tilde{\alpha}} - s
\sum_{a=1}^s\langle F_s \sigma_{i_0}^a\sigma_{i_0}^{s+1}
\rangle_{t,\tilde{\alpha}} + \frac{s(s+1)}{2!}\langle F_s
\sigma_{i_0}^{s+1}\sigma_{i_0}^{s+2} \rangle_{t,\tilde{\alpha}}]
\quad
\\ \nonumber
&+& \quad 2\tilde{\alpha}\theta^3 [\sum_{a<b<c}^{1,s}\langle F_s
\sigma_{i_0}^a\sigma_{i_0}^b\sigma_{i_0}^c
\rangle_{t,\tilde{\alpha}} - s \sum_{a<b}^{1,s}\langle F_s
\sigma_{i_0}^a\sigma_{i_0}^b\sigma_{i_0}^{s+1}
\rangle_{t,\tilde{\alpha}}
\\ \nonumber
&+& \frac{s(s+1)}{2!}\sum_{a=1}^{s}\langle F_s \sigma_{i_0}^a
\sigma_{i_0}^{s+1}\sigma_{i_0}^{s+2}
\rangle_{t,\tilde{\alpha}}\quad + \frac{s(s+1)(s+2)}{3!} \langle
F_s \sigma_{i_0}^{s+1}\sigma_{i_0}^{s+2}\sigma_{i_0}^{s+3}
\rangle_{t,\tilde{\alpha}}] \,
\end{eqnarray}
where we neglected terms $O(\theta^3)$.
\end{proposition}
\bigskip
\textbf{Proof}
\newline
The proof works by direct calculation:
\begin{eqnarray}
\frac{\partial\langle F_s \rangle_{t,\tilde{\alpha}}}{\partial t}
&=& \frac{\partial}{\partial t} \mathbb{E} \Big[
\frac{\sum_{\{\sigma\}}F_s
e^{\sum_{a=1}^s(\beta\sum_{\nu=1}^{P_{\tilde{\alpha} N}}
\sigma_{i_\nu}^a\sigma_{j_\nu}^a + \beta
\sum_{\nu=1}^{P_{2\tilde{\alpha }t}} \sigma_{i_\nu}^a)}}
{\sum_{\{\sigma\}}
e^{\sum_{a=1}^s(\beta\sum_{\nu=1}^{P_{\tilde{\alpha} N}}
\sigma_{i_\nu}^a\sigma_{j_\nu}^a + \beta
\sum_{\nu=1}^{P_{2\tilde{\alpha }t}} \sigma_{i_\nu}^a)}}\Big]
\\ \nonumber
&& = 2\tilde{\alpha} \mathbb{E} \Big[ \frac{\sum_{\{\sigma\}}F_s
e^{\sum_{a=1}^s(\beta\sigma_{i_0}^a +
\beta\sum_{\nu=1}^{P_{\tilde{\alpha} N}}
\sigma_{i_\nu}^a\sigma_{j_\nu}^a + \beta
\sum_{\nu=1}^{P_{2\tilde{\alpha }t}} \sigma_{i_\nu}^a)}}
{\sum_{\{\sigma\}} e^{\sum_{a=1}^s(\beta\sigma_{i_0}^a +
\beta\sum_{\nu=1}^{P_{\tilde{\alpha} N}}
\sigma_{i_\nu}^a\sigma_{j_\nu}^a + \beta
\sum_{\nu=1}^{P_{2\tilde{\alpha }t}} \sigma_{i_\nu}^a)}}\Big] -
2\tilde{\alpha}\langle F_s \rangle_{t,\tilde{\alpha}}
\\ \nonumber
&& = 2\tilde{\alpha} \mathbb{E}\Big[ \frac{\tilde{\Omega}_t (F_s
e^{\sum_{a=1}^s\beta\sigma_{i_0}^a})} {\tilde{\Omega}_t
(e^{\sum_{a=1}^s\beta\sigma_{i_0}^a})}\Big] -
2\tilde{\alpha}\langle F_s \rangle_{t,\tilde{\alpha}}
\\ \nonumber
&& = 2\tilde{\alpha} \mathbb{E}\Big[ \frac{\tilde{\Omega}_t (F_s
\Pi_{a=1}^{s}(\cosh\beta + \sigma_{i_0}^a\sinh\beta))}
{\tilde{\Omega}_t (\Pi_{a=1}^{s}(\cosh\beta +
\sigma_{i_0}^a\sinh\beta))}\Big] - 2\tilde{\alpha}\langle F_s
\rangle_{t,\tilde{\alpha}}
\\ \nonumber
&& = 2\tilde{\alpha} \mathbb{E}\Big[ \frac{\tilde{\Omega}_t (F_s
\Pi_{a=1}^{s}(1 + \sigma_{i_0}^a\theta))} {(1 +
\tilde{\omega}_t(\sigma_{i_0}^a)\theta)^s}\Big] -
2\tilde{\alpha}\langle F_s \rangle_{t,\tilde{\alpha}} =
\end{eqnarray}
\medskip
Now noting that
\begin{eqnarray}
\Pi_{a=1}^{s}(1 + \sigma_{i_0}^a\theta) &=& 1 +
\sum_{a=1}^{s}\sigma_{i_0}^a\theta +
\sum_{a<b}^{1,s}\sigma_{i_0}^a\sigma_{i_0}^b\theta^2 +
\sum_{a<b<c}^{1,s}\sigma_{i_0}^a\sigma_{i_0}^b\sigma_{i_0}^c\theta^3
+ ...
\\ \nonumber
\frac{1}{(1 + \tilde{\omega}_t \theta)^s} &=& 1 -
s\tilde{\omega}_t \theta + \frac{s(s+1)}{2!}\tilde{\omega}_t^2
\theta^2 - \frac{s(s+1)(s+2)}{3!}\tilde{\omega}_t^3 \theta^3 + ...
\end{eqnarray}
\medskip
we obtain
\begin{eqnarray}
\frac{\partial\langle F_s \rangle_{t,\tilde{\alpha}}}{\partial t}
= 2\tilde{\alpha} \mathbb{E}\Big[ \tilde{\Omega}_t \Big(F_s(1 +
\sum_{a=1}^{s}\sigma_{i_0}^a\theta +
\sum_{a<b}^{1,s}\sigma_{i_0}^a\sigma_{i_0}^b\theta^2 +
\sum_{a<b<c}^{1,s}\sigma_{i_0}^a\sigma_{i_0}^b\sigma_{i_0}^c\theta^3
+ ...)\Big) \times
\\ \nonumber
\times \Big(1 - s\tilde{\omega}_t \theta +
\frac{s(s+1)}{2!}\tilde{\omega}_t^2 \theta^2 -
\frac{s(s+1)(s+2)}{3!}\tilde{\omega}_t^3 \theta^3 + ...\Big)\Big]
- 2\tilde{\alpha}\langle F_s \rangle_{t,\tilde{\alpha}}.
\end{eqnarray}
from which our thesis follows $\Box$.

\section{Free energy analysis}\label{laC}

Now that we exploited the machinery we can start applying it to
the free energy. Let us at first work out its streaming with
respect to the plan $(\alpha,\beta)$:

\begin{eqnarray}
&& \frac{\partial A(\alpha,\beta)}{\partial\beta} = -
\frac{\langle H \rangle}{N} = \frac{1}{N}\mathbb{E}
\Big(\frac{1}{Z_N} \sum_{\{\sigma\}} \sum_{\nu=1}^{P_{\alpha N}}
\sigma_{i_\nu}\sigma_{j_\nu} e^{-\beta H_N(\alpha)}\Big)
\\ \nonumber
&&=\frac{1}{N}\sum_{k=1}^{\infty} k \pi(k-1,\alpha N) \mathbb{E}
[\omega(\sigma_{i_k}\sigma_{j_k})_k] = \alpha \sum_{k=1}^{\infty}
\pi(k-1,\alpha N) \mathbb{E}
\Big[\frac{\omega(\sigma_{i_k}\sigma_{j_k}e^{\beta\sigma_{i_k}\sigma_{j_k}})_{k-1}}
{\omega(e^{\beta\sigma_{i_k}\sigma_{j_k}})_{k-1}}\Big]
\\ \nonumber
&&= \alpha \mathbb{E}
\Big[\frac{\omega(\sigma_{i_k}\sigma_{j_k}(\cosh\beta +
\sigma_{i_k}\sigma_{j_k}\sinh\beta))} {\omega(\cosh\beta +
\sigma_{i_k}\sigma_{j_k}\sinh\beta)}\Big] = \alpha
\mathbb{E}\Big[\frac{\omega(\sigma_{i_k}\sigma_{j_k}) + \theta} {1
+ \omega(\sigma_{i_k}\sigma_{j_k})\theta}\Big]
\end{eqnarray}
\medskip
by which we get (and with similar calculations for
$\partial_{\alpha}A(\alpha,\beta)$ that we omit for the sake of
simplicity):
\begin{eqnarray}
\frac{\partial A(\alpha,\beta)}{\partial\beta} &=& \alpha\theta -
\alpha \sum_{n=1}^{\infty} (-1)^n (1 - \theta^2) \theta^{n-1}
\langle q_{1,..,n}^2 \rangle
\\ \label{dAda} \frac{\partial A(\alpha,\beta)}{\partial\alpha} &=& \ln
\cosh\beta - \sum_{n=1}^{\infty} \frac{(-1)^n}{n} \theta^{n}
\langle q_{1,..,n}^2 \rangle
\end{eqnarray}
\bigskip
Now remembering Theorem (\ref{main}) and assuming critical
behavior (that we will verify a fortiori in sec. (\ref{critico}))
we move for a different formulation of the free energy by
considering the cavity function as the integral of its derivative.
In a nutshell the idea is as follows: Due to the second order
nature of the phase transition for this model (i.e. criticality
that so far is assumed) we can expand the free energy in terms of
the whole series of order parameters. Of course it is impossible
to manage all these infinite overlap correlation functions to get
a full solution of the model in the whole ($\alpha,\beta$) plane
but it is possible to show by rigorous bounds that close to the
critical line (that we are going to find soon) higher order
overlaps scale with higher order critical exponents so we are
allowed to neglect higher orders close to this line and we can
investigate deeply criticality, which is the topic of the paper.
\newline
To this task let us expand the cavity functions as
\medskip
\begin{eqnarray}
\Psi(\tilde{\alpha},\beta,t) &=& \int_0^t \frac{\partial
\Psi}{\partial t'}dt'
\\ \nonumber
&=& 2\tilde{\alpha}t \log\cosh\beta + \tilde{\beta}\int_0^t\langle
m \rangle_{t',\tilde{\alpha}}dt' -
\frac{1}{2}\tilde{\beta}\theta\int_0^t\langle q_{12}
\rangle_{t',\tilde{\alpha}}dt' + O(\theta^3)
\end{eqnarray}
where $\tilde{\beta} = 2\tilde{\alpha}\theta \rightarrow \beta' =
2\alpha\theta$ for $N\rightarrow \infty$. Now using the streaming
equation as dictated by Proposition (\ref{stream}) we can write
the overlaps appearing in the expression of $\Psi$ as polynomials
of higher order filled overlaps so to obtain a straightforward
polynomial back-integration for the $\Psi$ as they no longer will
depend on the interpolating parameter thanks to Theorem
(\ref{saturi}).
\newline
For the sake of simplicity the $\tilde{\alpha}$-dependence of the
overlaps will be omitted keeping in mind that our results are all
taken in the thermodynamic limit and so we can quietly exchange
$\tilde{\alpha}$ with $\alpha$ in these passages.
\medskip
The first equation we deal with is:
\begin{equation}
\frac{d\langle m \rangle_t}{dt} =
\tilde{\beta}[\langle m^2 \rangle - \langle m_1m_2 \rangle_t]
\end{equation}
\medskip
where $\langle m_1m_2 \rangle $ is not filled and so we have to go
further in the procedure and derive it in order to obtain filled
monomials:
\begin{equation}\label{m1m2}
\frac{d\langle m_1m_2 \rangle_t}{dt} = 2\tilde{\beta}[\langle
m_1^2m_2 \rangle_t - \langle m_1m_2m_3 \rangle_t] +
\tilde{\beta}\theta [\langle m_1m_2q_{12} \rangle - 4\langle
m_1m_2q_{13} \rangle_t + 3\langle m_1m_2q_{34} \rangle_t].
\end{equation}
In this expression we stress the presence of the filled overlap
$\langle m_1m_2q_{12} \rangle$ and of $\langle m_1^2m_2 \rangle_t$
which can be saturated in just one derivation. Wishing to have an
expansion for $\langle m \rangle_t$  up to the third order in
$\theta$, it is easy to check that the saturation of the other
overlaps in the last derivative would carry terms of higher order
and so we can stop the procedure at the next step
\begin{equation}
\frac{d\langle m_1^2m_2 \rangle_t}{dt} = \tilde{\beta}[\langle
m_1^2m_2^2 \rangle] + \tilde{\beta}[\mbox{unfilled terms}] +
O(\theta^2)
\end{equation}
from which integrating back in $t$
\begin{equation}
\langle m_1^2m_2 \rangle_t = \tilde{\beta}[\langle m_1^2m_2^2
\rangle]t
\end{equation}
inserting now this result in the expression (\ref{m1m2}) and
integrating again in $t$ we find
\begin{equation}
\langle m_1m_2 \rangle_t =
\tilde{\beta}\theta\langle m_1m_2q_{12} \rangle t +
\tilde{\beta}^2\langle m_1^2m_2^2 \rangle t^2
\end{equation}
and coming back to $\langle m \rangle_t$ we get
\medskip
\begin{equation}\label{magneto}
\quad \langle m \rangle_t = \tilde{\beta}\langle m^2 \rangle t -
\frac{\tilde{\beta}^2\theta}{2}\langle m_1m_2q_{12} \rangle t^2 -
\frac{\tilde{\beta}^3}{3}\langle m_1^2m_2^2 \rangle t^3
\end{equation}
which is the attempted result. Let us move our attention to
$\langle q_{12}\rangle_t$, analogously we can write
\begin{equation}
\frac{d\langle q_{12} \rangle_t}{dt} =
2\tilde{\beta}[\langle m_1q_{12} \rangle_t - \langle m_3q_{12} \rangle_t] +
\tilde{\beta}\theta [\langle q_{12}^2 \rangle - 4\langle q_{12}q_{13} \rangle_t +
3\langle q_{12}q_{34} \rangle_t]
\end{equation}
and consequently obtain
\begin{equation}\label{overlap}
\quad \langle q_{12} \rangle_t = \tilde{\beta}\theta \langle
q_{12}^2 \rangle t + \tilde{\beta}^2\langle m_1m_2q_{12} \rangle
t^2 + O(\theta^4).
\end{equation}
With the two expansion above, in the $N\rightarrow\infty$ limit,
putting $t=1$ we have
\medskip
\begin{equation}
\Psi(\alpha,\beta,t=1) = 2\alpha \ln\cosh\beta +
\frac{\beta'}{2}\langle m^2 \rangle - \frac{\beta'^4}{12}\langle
m_1^2m_2^2 \rangle - \frac{\beta'^2\theta^2}{4}\langle q_{12}^2
\rangle - \frac{\beta'^3\theta}{3}\langle m_1m_2q_{12} \rangle +
O(\theta^6)
\end{equation}
\medskip
At this point we have all the ingredients to write down the
polynomial expansion for the free energy function as stated in the
next:
\medskip
\begin{proposition}
A general expansion via stochastically stable terms for the free
energy of the diluted Ising model can be written as
\begin{eqnarray}\label{A}
A(\alpha,\beta) &=& \ln2 + \alpha \ln\cosh\beta  +
\frac{\beta'}{2}\left(\beta' - 1\right)\langle m_1^2 \rangle  +
\\ \nonumber
&-&  \frac{\beta'^4}{12}\langle m_1^2m_2^2 \rangle  -
\frac{\beta'^2}{8\alpha}\left(\frac{\beta'^2}{2\alpha} -
1\right)\langle q_{12}^2 \rangle  -
\frac{\beta'^4}{6\alpha}\langle m_1m_2q_{12} \rangle  +
O(\theta^6).
\end{eqnarray}
\end{proposition}
It is immediate to check that the above expression, in the ergodic
region where the averages of all the order parameters vanish,
reduces to the well known high-temperature (or high connectivity)
solution \cite{dsg} (i.e. $A(\alpha,\beta)=\ln 2 + \alpha
\log\cosh \beta$).
\newline
Of course we are neglecting $\theta^6$ higher order terms because
we are interested in an expansion holding close to the critical
line, but we are not allowed to truncate the series for a general
point in the phase space far beyond the ergodic region.

\section{Critical behavior}\label{critico}

Now we want to analyze the critical behavior of the model: we find
the critical line where the ergodicity breaks, we obtain the
critical exponent of the magnetization and the susceptibility and
at the end we show that within our framework the lacking of
ergodicity can be explained as the breaking of commutativity of
the infinite volume limit against our cavity field, thought of as
a properly chosen field, vanishing in the thermodynamic limit too,
accordingly with the standard prescription of statistical
mechanics \cite{amit}.

\subsection{Critical line}

Let us firstly define the rescaled magnetization $\xi_N$ as $\xi_N
= \sqrt{N}m_N$. By applying the gauge transformation $\sigma_i
\rightarrow \sigma_i\sigma_{N+1}$ in the expression for the
quenched average of the magnetization (eq. (\ref{magneto})) and
multiplying it times $N$ so to switch to $\xi^2_N$, setting $t=1$
and sending $N \rightarrow \infty$ we obtain
\medskip
\begin{equation}\label{m2}
\langle \xi_1^2 \rangle = \frac{\beta'^3}{3(\beta'-1)}\langle
\xi_1\xi_2 m_1 m_2 \rangle +
\frac{\beta'^2\theta}{2(\beta'-1)}\langle \xi_1\xi_2 q_{12}
\rangle + O(\frac{\theta^5}{\beta'-1})
\end{equation}
\medskip
by which we see (again remembering criticality and so forgetting
higher order terms) that the only possible divergence of the
(centered and rescaled) fluctuations of the magnetization happens
at the value $\beta' = 1$ which gives $2\alpha\theta=1$ as the
critical line, in perfect agreement with \cite{dsg}.
\medskip
The same critical line can be found  easier simply by looking at
the expression (\ref{A}) as follows: remembering  that in the
ergodic phase the minimum of the free energy corresponds to a zero
order parameter (i.e.$\sqrt{\langle m^2 \rangle}=0$), this implies
the coefficient of the second order $a(\beta') =
\frac{\beta'}{2}(\beta' - 1)$ to be positive. Anyway immediately
below the critical line values of the magnetization different from
zero must be allowed (by definition otherwise we were not crossing
a critical line) and this can be possible if and only if
$a(\beta') \leq 0$. Consequently (and using once more the second
order nature of the transition) on the critical line we must have
$a(\beta') = 0$ and this gives again $2\alpha\theta = 1$.
\begin{figure}[tb]\bc
\includegraphics[angle=0,width=6cm,height=50mm]
{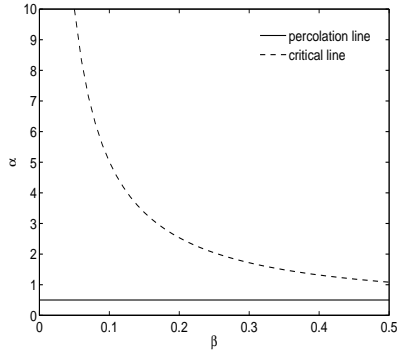} \caption{\label{phase} Phase diagram: below
$\alpha_c=0.5$ there is no giant component in the Erdos-Renyi
graph, $\alpha_c$ defines the percolation threshold. Above at left
of the critical line the system behaves ergodically, conversely on
the right ergodicity is broken and the system displays
magnetization.} \ec
\end{figure}

\subsection{Critical exponents and bounds}

Now let us move to the critical exponents:
\newline
Critical exponents are needed to characterize singularities of the
theory at the critical line and, for us, these indexes are the
ones related to the magnetization $\langle m \rangle$ and to the
susceptibility $\langle \chi \rangle$.
\newline
We define $\tau=(2\alpha \tanh \beta-1)$ and we write $\langle
m(\tau) \rangle \sim G_0 \cdot \tau^{\delta}$ and $\langle
\chi(\tau) \rangle \sim G_0 \cdot \tau^{\gamma}$, where the symbol
$\sim$ has the meaning that the term at the second member is the
dominant but there are corrections of higher order.
\newline
Remembering the expansion of the squared magnetization that we
rewrite for completeness
\begin{equation}\label{m2}
\langle m^2 \rangle =
\frac{\beta'^3}{3(\beta'-1)}\langle m_1^2m_2^2 \rangle +
\frac{\beta'^2\theta}{2(\beta'-1)}\langle m_1m_2q_{12} \rangle + O(\frac{\theta^5}{\beta'-1})
\end{equation}
and considering that using the same gauge transformation $\sigma_i
\rightarrow \sigma_i\sigma_{N+1}$ on (eq.(\ref{overlap})) we have
for the two replica overlap the following representation
\begin{equation}
\langle q_{12}^2 \rangle = - \frac{\beta'^2}{(\beta'\theta -
1)}\langle m_1m_2q_{12} \rangle + O(\theta^6)
\end{equation}
we can arrive by simple algebraic calculations to write down the
free energy, close to the critical line of course, depending only
by the two parameters $\langle m^2 \rangle$ and $\langle q_{12}^2
\rangle$
\medskip
\begin{equation}\label{38}
A(\alpha,\beta) = \ln2 + \alpha \ln \cosh\beta +
\frac{\beta'}{4}\left(\beta' - 1\right)\langle m_1^2 \rangle -
\frac{\beta'^2}{48\alpha}\left(\frac{\beta'^2}{2\alpha} -
1\right)\langle q_{12}^2 \rangle + O(\theta^6)
\end{equation}
\medskip
\newline
By a comparison of the formula obtained by deriving
$A(\alpha,\beta)$ as expressed by eq.(\ref{38}) and the expression
we have previously found (eq. (\ref{dAda})) that we report for the
sake of readability,
\begin{equation}\label{dAda}
\frac{\partial A(\alpha,\beta)}{\partial\alpha} =
\ln\cosh\beta - \sum_{n=1}^{\infty}
\frac{(-1)^n}{n} \theta^{n}
\langle q_{1,..,n}^2 \rangle
\end{equation}
it is immediate to see that we have
\medskip
\begin{equation}
\frac{\partial}{\partial\alpha}\Big[\frac{\beta'}{4}(\beta' -
1)\langle m_1^2 \rangle\Big]= \theta \langle m_1^2 \rangle.
\end{equation}
\medskip
If we put ourselves close to the value $\beta'=1$ and make a
change of variable $\tau = \beta' - 1$ with $\partial_{\alpha} =
2\theta
\partial_{\tau}$ we get
\begin{equation}
\frac{\partial}{\partial\alpha}\Big[\frac{\beta'}{4}(\beta' -
1)\langle m_1^2 \rangle\Big] \sim
\frac{\theta}{2}\frac{\partial}{\partial\tau}[\tau\langle m_1^2
\rangle] = \frac{\theta}{2}\langle m_1^2 \rangle +
\frac{\theta\tau}{2}\frac{\partial\langle m_1^2 \rangle}{\partial
\tau} = \theta \langle m_1^2 \rangle
\end{equation}
by which we easily obtain
\begin{equation}
\frac{\partial\langle m_1^2 \rangle}{\langle m_1^2 \rangle} =
\frac{\partial \tau}{\tau} \qquad \Rightarrow \langle m_1^2
\rangle \sim \tau \qquad \Rightarrow \sqrt{\langle m_1^2 \rangle}
\sim \tau^{\frac{1}{2}}
\end{equation}
Therefore we get that the critical exponent for the magnetization,
$\delta = 1/2$, which turns out to be the same as in the fully
connected counterpart \cite{abarra}\cite{ellis}, in agreement with
the disordered extension of this model \cite{ barra5}.
\newline
Again, by simple direct calculations, once we get the critical
exponent for the magnetization it is straightforward to show that
the susceptibility $\langle \chi \rangle$ \cite{amit} obeys \be
\langle \chi \rangle \sim |\tau|^{-1} \ee close to the critical
line, by which we find its critical exponent to be once again in
agreement with the classical fully connected counterpart
\cite{amit}.

\bigskip

Now we want to show some wrong results which a naive calculation
would suggest so to emphasize the importance of the bounds
 relating different monomials that we are going to discuss immediately later
 \cite{barra6}\cite{talax}. Then in  the next subsection,
 we explain what is the physics behind this picture by providing a mechanism for the breaking of the ergodicity.
\newline
The point on which we focus is the following: if we wish to
perform the same procedure we performed on $\langle m^2 \rangle$,
applying blindly saturability below to the first critical line, to
the $2$ replica overlap $\langle q_{12}^2 \rangle$ we would gain
\medskip
\begin{equation}
\sqrt{\langle q_{12}^2 \rangle} \,\sim \, \tau_2 = (\beta'\theta - 1)
\end{equation}
identifying $\theta_{c_2} = 1/(2\alpha)^{1/2}$ as another critical
temperature, or better, the critical temperature typical of
$\langle q_{12}^2 \rangle$. In the same way we could find
$\theta_{c_n}$ for every $q_{1...,n}^2$ obtaining
$$
\theta_{c_n} = 1/(2\alpha)^{1/n}
$$
such that, at the end, we obtain a scenario with several
transition lines, one for every order parameter.
\newline
This is not a possible scenario, as generally explained for
instance in \cite{abbarra} and as dictated, in this model, by the
following
\begin{proposition}
As soon as the first order parameter (the magnetization) starts taking
values different from zero, the same happens to all the other
order parameters
\begin{equation}
\langle q_{1,..,n}^2 \rangle = \mathbb{E}_k \mathbb{E}  _i
[\omega^n(\sigma_{i_1}\sigma_{i_2})] \geq (E_k E_i
[\omega(\sigma_{i_1}\sigma_{i_2})])^n = \langle m^2 \rangle^n
\quad \forall n
\end{equation}
\end{proposition}
We omit the proof details as they are a simple application of
Jensen inequality.

\subsection{Saturability breaking}

So far we showed that it is not possible to have several
transition lines, one for every order parameter. Now we want to
understand why there is just one critical line by applying the
theory developed in \cite{barra6} to this model.
\newline
Starting from Theorem (\ref{main}) that we recall for simplicity
\begin{equation}
A(\alpha,\beta) = \ln2 - \alpha \frac{\partial
A(\alpha,\beta)}{\partial\alpha} + \Psi(\alpha,\beta,t=1)
\end{equation}
we want to show the phase transition expressed by the
non-commutativity among the thermodynamic limit and the vanishing
perturbation.
\newline
Again for simplicity we report the expansion of $A(\alpha,\beta)$
we have previously built
\begin{equation}\label{Afinale}
A(\alpha,\beta) = \ln2 + \alpha \ln \cosh\beta +
\frac{\beta'}{4}\left(\beta' - 1\right)\langle m_1^2 \rangle -
\frac{\beta'^2}{48\alpha}\left(\frac{\beta'^2}{2\alpha} -
1\right)\langle q_{12}^2 \rangle + O(\theta^6)
\end{equation}
that we obtained by considering the cavity function as the
integral of its $t$-derivative
\begin{equation}\label{cavita}
\Psi(\tilde{\alpha},t) =
\int_o^t \frac{\partial\Psi(\tilde{\alpha},t')}{\partial t'}dt' =
\int_o^t2\tilde{\alpha}\ln\cosh\beta dt' -
2\tilde{\alpha}\sum_{n=1}^{\infty}\frac{(-1)^n}{n}
\theta^n\int_o^t\langle q_{1,...,n} \rangle_t'dt'
\end{equation}
and performing, via the streaming equation, a saturating procedure
of consecutive $t$-derivatives upon the overlaps in order to
express them as functions of higher order filled terms. Then the
only thing we had to do was sending $N$ to infinity, carrying out
of the integral the overlaps and then putting $t=1$ to evaluate
$\Psi(\alpha,\beta,t=1)$. The result of this procedure brings to
equation (\ref{Afinale}).
\newline
But what if we exchanged the limit order by sending
$t\rightarrow1$ first and taking the thermodynamic limit after?
\newline
It is easy to note that all the overlaps appearing in
eq.(\ref{cavita}) are fillable such that we can avoid the
saturation procedure simply by setting $t=1$ first and then
sending N to infinity. In this way, thanks to theorem
(\ref{saturabili}), each fillable overlap is transformed in a
filled $t$-independent one and this kills all the correlations
among different replicas and allows us to write
\begin{equation}
A(\alpha,\beta) = \ln2 + \alpha\ln\cosh\beta
- \sum_{n=1}^{\infty}
\frac{(-1)^n}{n} \theta^{n}
\langle q_{1,..,n}^2 \rangle =
\ln2 + \alpha \frac{\partial A(\alpha,\beta)}{\partial\alpha}
\end{equation}
being clearly
\begin{eqnarray}
\lim_{N\rightarrow\infty}\lim_{t\rightarrow1}\Psi(\tilde{\alpha},\beta,t)
&=& \lim_{N\rightarrow\infty}\lim_{t\rightarrow1} \int_o^t
\frac{\partial\Psi(\tilde{\alpha},\beta,t')}{\partial t'}dt'
\\ \nonumber
&=& \lim_{N\rightarrow\infty}\lim_{t\rightarrow1}
[ \int_o^t2\tilde{\alpha}\log\cosh\beta dt' -
2\tilde{\alpha}\sum_{n=1}^{\infty}\frac{(-1)^n}{n}
\theta^n\int_o^t\langle q_{1,...,n} \rangle_t'dt' ]
\\ \nonumber
&=& 2\alpha\ln\cosh\beta - 2\alpha \sum_{n=1}^{\infty}
\frac{(-1)^n}{n} \theta^{n} \langle q_{1,..,n}^2 \rangle = 2\alpha
\frac{\partial A(\alpha,\beta)}{\partial\alpha}.
\end{eqnarray}
\newline
In particular, retaining just the first two terms of the
expansions, we show the difference between the two results.
\begin{itemize}
\item $\lim_{t\rightarrow1} \lim_{N\rightarrow\infty}$

\begin{equation}\label{51}
A(\alpha,\beta) = \ln2 + \alpha\ln\cosh\beta +
\frac{\beta'}{4}(\beta' - 1)\langle m_1^2 \rangle + O(\theta^3)
\end{equation}

\item $\lim_{N\rightarrow\infty} \lim_{t\rightarrow1}$

\begin{equation}\label{52}
A(\alpha,\beta) = \ln2 + \alpha\ln\cosh\beta +
\frac{\beta'}{2}\langle m_1^2 \rangle + O(\theta^3)
\end{equation}
\end{itemize}
We immediately recognize that eq.(\ref{51}) is the correct
expression holding also below the critical line. When the system
lives in the ergodic region all the order parameters are zero and
it reduces to $\alpha\ln\cosh(\beta)$ which is the ergodic
solution, if we cross the critical line the formula takes into
account the phase transition encoded in the coefficient of the
second order and gives the correct expression immediately below.
\newline It is also straightforward to recognize as the ergodic
solution eq.(\ref{52}) which can be the correct one only above the
critical line \cite{dsg}.
\newline
At the end we saw that there exists only and only one critical
line for all the order parameters. We saw that this line can be
depicted as the breaking of commutativity among the  infinite
limit operation and the setting of $t=1$ (relaxing the Boltzmann
factor from the interpolant avoiding the trivial way $t=0$).
Hence, which can be the genesis of the transition for all the
other order parameters? The correlations among them and the
magnetization, as clearly put in evidence in the free energy
expansion (see eq.(\ref{A})). And lastly, which is the origin of
these correlation? The {\em saturability property} that the model
exhibits as stated by Theorem (\ref{saturabili}). In fact, at the
critical line (which is the last line, from above, where gauge
invariance is a symmetry of the Boltzmann state too, thanks to the
continuity of the transition) saturability easily shows that \be
\lim_{N \rightarrow \infty} \langle m_1 q_{12} \rangle_t = \langle
 m_1 m_2 q_{12} \rangle,\ee
explaining the birth of the correlations among the various order
parameters (we reported just the first two as an example).

\subsection{Self-averaging properties}\label{selfa}

We have previously  shown how filled overlaps become
asymptotically $t$-independent  when N grows to infinity. Starting
from this we can find identities stating the self-averaging of the
order parameters as $\langle m^2 \rangle$.
\newline
In particular we are going to take these overlaps and calculate
their derivative with respect to $t$; then by applying the gauge
transformation and setting $t=1, N\rightarrow \infty$ we can write
down the self-averaging relations.
\medskip
\begin{equation}
0=\partial_{t}\langle m_1^2 \rangle_t = \beta'[\langle m_1^3
\rangle_t - \langle m_1^2m_2 \rangle_t] \Rightarrow [\langle m_1^4
\rangle - \langle m_1^2m_2^2 \rangle] = \mathbb{E}[\Omega(m^4) -
\Omega^2(m^2)].
\end{equation}
Coherently with \cite{dsg} we find standard self-averaging for the
magnetization.
\newline
More interesting is the situation concerning the overlap but we
actually lack a complete mathematical control.
\newline
By applying the same trick as before we can equate to zero (in the
large $N$ limit) the $t$-derivative of the squared overlap (which
is stochastically stable), consequently we get two terms.
\medskip
\begin{eqnarray}
0=\partial_{t}\langle q_{12}^2 \rangle_t &=& 2\beta'[\langle
m_1q_{12}^2 \rangle_t - \langle m_3q_{12}^2 \rangle_t] +
\beta'\theta[\langle q_{12}^3 \rangle_t - 4\langle
q_{12}^2q_{13}\rangle_t + 3\langle q_{12}^2q_{34} \rangle_t]
\\ \label{ipotesi}
&\Rightarrow& [\langle m_1^2q_{12}^2 \rangle - \langle
m_3^2q_{12}^2 \rangle] = 0 \qquad [\langle q_{12}^4 \rangle -
4\langle q_{12}^2q_{13}^2\rangle + 3\langle q_{12}^2q_{34}^2
\rangle] = 0
\end{eqnarray}
A-priori we can not assume factorization of the series so to put
to zero each term separately (as we did in eq.(\ref{ipotesi})),
however, close to the critical line surely the second term is an
higher order and we can reasonably set to zero the first.
Furthermore as the second term at the r.h.s. has a pre-factor
$\propto \alpha^{-1}$ differing from the first term it would be
difficult to imagine the opposite.
\newline
It is in fact very natural to assume that the two terms can be set
to zero separately in the whole $(\alpha,\beta)$ plane and this is
very interesting because the second term is a very well known
relation in the field of spin glasses
 \cite{ggi}\cite{guerra2}\cite{ac}\cite{gg}\cite{barra1} suggesting
a common structure among different kinds of disordered systems,
the only sharing feature among diluted ferromagnets and
spin-glasses being some kind of disorder (topological in the
former, frustrating in the latter).
\newline
We are not going to deepen this point as it is under investigation
in \cite{CCN} where the same set of relations (and more) is found
and discussed.

\section{Numerics}\label{numerics}

In this section we analyze, from the numerical point of view, the
ferromagnetic system previously introduced by performing extensive
Monte Carlo simulations with the Metropolis algorithm
\cite{barkema}.
\newline
The Erdos-Renyi random graph is constructed by taking $N$ sites
and introducing a bond between each pair of sites with probability
$p=\bar{\alpha}/(N-1)$, in such a way that the average
connectivity per node is just $\bar{\alpha}$. Clearly, when $p=1$
the complete graph is recovered.
\newline
The simplest version of the diluted Curie-Weiss Hamiltonian has a
Poisson variable per bond as
$H_N=-\sum_{ij}\sum_{\nu=0}^{P_{\bar{\alpha}/
N}}\sigma_{i_{\nu}}\sigma_{j_{\nu}}$, and it is the easiest
approach when dealing with numerics.
\newline
For the analytical investigation we choose a slightly changed
version (see eq.(\ref{ham})): each link gets a bond with
probability close to $\alpha/N$ for large $N$; the probabilities
of getting two, three bonds scale as $1/N^2,1/N^3$ therefore
negligible  in the thermodynamic limit.
\newline
Working with directed links (as we do in the analytical framework)
the probability of having a bond on any undirected link is twice
the probability for directed link (i.e. $2\alpha/N$). Hence, for
large $N$, each site has average connectivity $2\alpha$. Finally
in this way we allow self-loop but they add just
$\sigma$-independent constant to $H_N$ and are irrelevant, but we
take the advantage of dealing with an $H_N$ which is the sum of
independent identically distributed random variables, that is
useful for analytical investigation.
\newline
When comparing with numerics consequently we must keep in mind
that $\bar{\alpha} = 2 \alpha$.
\newline
In the simulation, once the network has been diluted, we place a
spin $\sigma_i$ on each node $i$ and allow it to interact with its
nearest-neighbors. Once the external parameter $\beta$ is fixed,
the system is driven by the single-spin dynamics and it eventually
relaxes to a stationary state characterized by well-defined
properties. More precisely, after a suitable time lapse $t_0$  and
for sufficiently large systems, measurements of a (specific)
physical observable $x(\sigma,\bar{\alpha},\beta)$ fluctuate
around an average value only depending on the external parameters
$\beta^{-1}$ and $\bar{\alpha}$.
\newline
Moreover, for a system $(\bar{\alpha},\beta)$ of a given finite size $N$ the extent of such fluctuations scales as $N^{-\frac{1}{2}}$
with the size of the system. The estimate of the thermodynamic
observables $\langle x \rangle$ is therefore obtained as an
average over a suitable number of (uncorrelated) measurements
performed when the system is reasonably close to the equilibrium
regime.
\newline
The estimate is further improved by averaging over different
realizations of the same system
$(\bar{\alpha},\beta)$. 
In summary,
$$
\langle x(\sigma,\bar{\alpha},\beta) \rangle = \mathbb{E} \left[
\frac{1}{M} \sum_{n=1}^{M} x(\sigma(t_n))\right] , \; \;
t_n=t_0+n\mathcal{T}
$$
where $\sigma(t)$ denotes the configuration of the magnetic system
at time step $t$ and $\mathcal{T}$ is the decorrelation parameter
(i.e. the time, in units of spin flips, needed to decorrelate a
given magnetic arrangement).
\newline
In general, statistical errors during a MC run in a given sample
result to be significantly smaller than those arising from the
ensemble averaging (see also \cite{szalma}).
\begin{figure}[tb]\bc
\includegraphics[height=50mm]{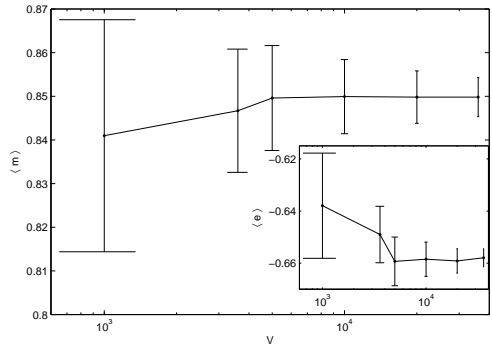}
\caption{\label{fig:fss}Finite size scaling for the  magnetization
 and the internal energy (inset) for $\bar{\alpha}=10$ and
$\frac{\beta}{\bar{\alpha}}=1.67$. All the measurements were
carried out in the stationary regime and the error bars represent
the fluctuations about the average values. We find good indication
of the convergence of the quantities on the size of the system and
thus of the existence of the thermodynamic limit.} \ec
\end{figure}
Figure (\ref{fig:fss}) shows the dependence of the macroscopic
observables $\langle m \rangle$ and $\langle e \rangle$ from the
size of the system; values are obtained starting from a
ferromagnetic arrangement, at the normalized inverse temperature
$\beta/\bar{\alpha}=1.67$. Notice that at this temperature the
system composed of $N=10^4$ is already very close to the
asymptotic regime. Analogous results are found for different
systems $(\bar{\alpha},\beta)$, with $\beta$ far enough from
 $\beta_c$.

\begin{figure}[tb]
\bc
\includegraphics[height=42mm]{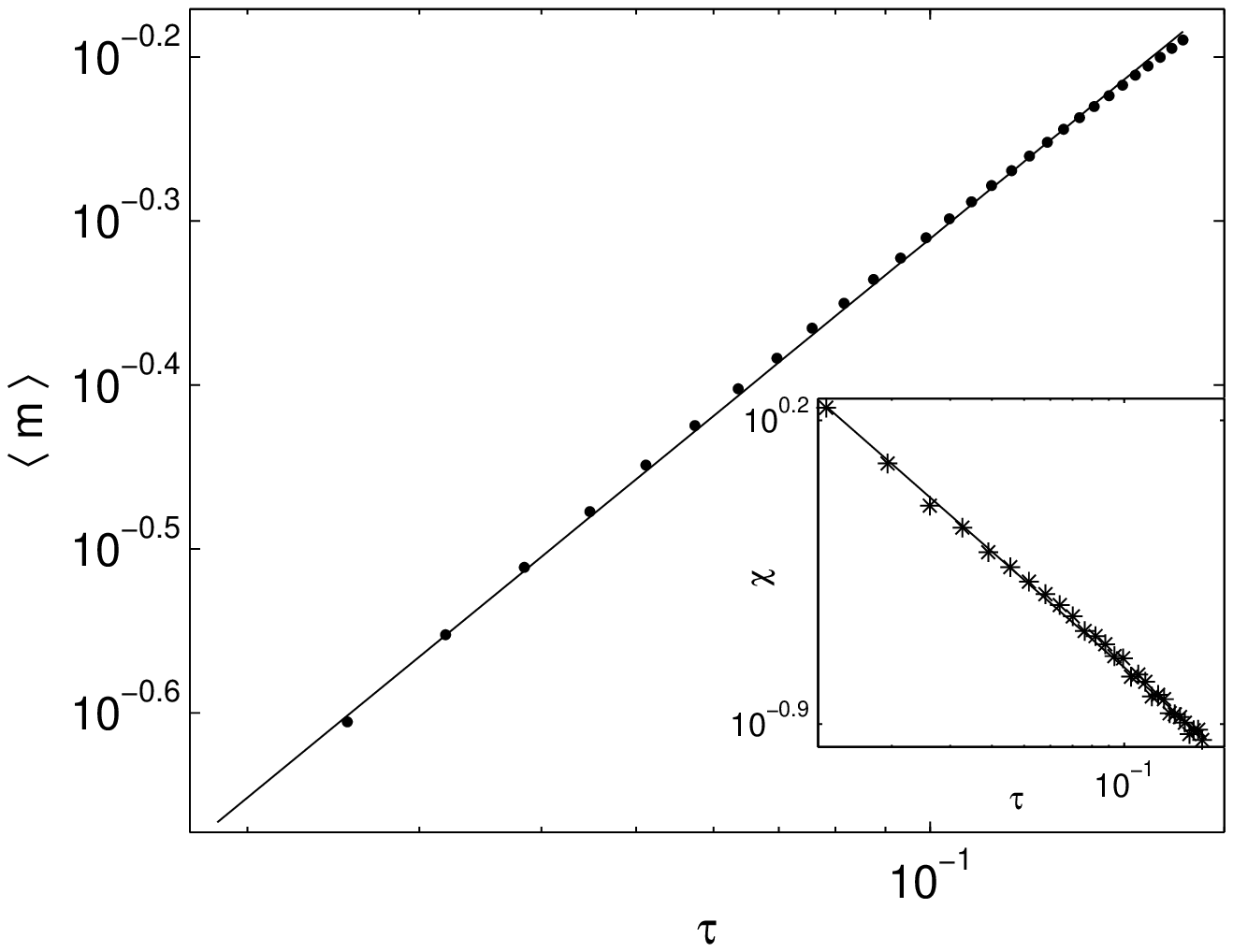}
\caption{\label{fig:crit} Log-log scale plot of magnetization
(main figure) and susceptibility (inset) versus the reduced
temperature $\tau=(|\beta-\beta_c|/\beta_c)^{-1}$ for
$\bar{\alpha} = 10$. Symbols represents data from numerical
simulations performed on systems of size $N=36000$, while lines
represent the best fit.
} \ec
\end{figure}

In the following we focus on systems of sufficiently large size so
to discard finite size effects. For a wide range of temperatures
and dilutions, we measure the average magnetization $\langle m
\rangle$ and energy $\langle e \rangle$, as well as the magnetic
susceptibility $\chi$, calculated as
$$
\chi(\beta,\bar{\alpha}) \equiv  \beta N \left[ \langle m^2
\rangle - \langle m \rangle^2 \right].
$$

Their profiles display the typical behavior expected for a
ferromagnet and, consistently with the theory, highlight a phase
transition at well defined temperatures $\beta_c(\alpha)$.
\newline
Now, we investigate in more detail the critical behavior of the
system. We collect accurate data of magnetization and
susceptibility, for different values of $\bar{\alpha}$ and for
temperatures approaching the critical one.
These data are used to estimate both the critical temperature and
the critical exponents for the
magnetization and susceptibility. In Fig.~(\ref{fig:crit}) we show
data as a function of the reduced temperature $\tau =
(|\beta-\beta_c|/\beta_c)^{-1}$ for $\bar{\alpha}=10$ and
$\bar{\alpha}=20$. The best fit for observables is
the power law
\begin{eqnarray}
\label{eq:powerlaw}
\langle m \rangle & \sim & \tau^{\delta},\ \beta > \beta_c \\
\chi & \sim & \tau^{\gamma}.
\end{eqnarray}
We obtain estimates for $\beta_c(\bar{\alpha})$,
$\delta(\bar{\alpha})$ and $\gamma(\bar{\alpha})$ by means of a
fitting procedure. Results are gathered in Tab.~\ref{tabexp}.
\begin{table}[htbp]\label{tabexp}
\begin{center}
\begin{tabular}{p{0.5cm}p{1cm}|p{0.5cm}p{1.2cm}p{1.2cm}p{1.2cm}}
\hline
& $\bar{\alpha}$ & & $\;\beta^{-1}_c$ &  $\;\delta$ & $\;\gamma$ \\
\hline
& 10 && $9.93$    &  $0.48$ & $-0.97$ \\
& 20 && $19.92$   &  $0.49$ & $-1.04$ \\
& 30 && $29.98$   &  $0.48$ & $-1.04$ \\
& 40 && $39.59$   &  $0.50$ & $-1.02$ \\
\hline
\end{tabular}
\end{center}
\caption{Estimates for the critical temperature and the critical
exponents $\delta$ and $\gamma$ obtained by a fitting procedure on
data from numerical simulations concerning Ising systems of size
$N=36000$ and different dilutions (we stress that analytically we
 get $\delta = 0.5$ and $\gamma=-1$). Errors on temperatures are $<2\%$, while for exponents are
within $5 \%$.}
\end{table}
Within the errors ($\leq 2\%$ for $\beta_c$ and $ \leq 5 \%$ for
the exponents), estimates for different values of $\bar{\alpha}$
agree and they are also consistent with the analytical results
exposed in Sec. (\ref{critico})
\newline
We also checked the critical line for the ergodicity breaking,
again finding optimal agreement with the criticality investigated
by means of analytical tools.

\section{Conclusions}\label{conclusions}

In this paper we developed the {\em interpolating cavity field}
technique for the mean field diluted ferromagnet. Once the general
framework has been built we used it to analyze criticality: we
found analytically  the critical line and the critical exponent of
the magnetization, whose self-averaging is also proved.  We
present an argument to explain the transition from an ergodic
phase to a broken ergodicity phase by the breaking of
commutativity of two limits, volume and applied field, as dictated
by standard statistical mechanics. We furthermore showed the
existence of only one critical line where all the multi-overlaps
start taking positive values as soon as the magnetization becomes
different from zero.  We proved this both mathematically by a
rigorous bound and physically by a mechanism that generates strong
correlations among magnetization and overlaps at the (unique)
critical line: {\itshape saturability}.
\newline
At the end a detailed numerical analysis of the model is
presented: by sharp Monte Carlo simulations the convergence of the
energy density (and the magnetization) to its limit is
investigated obtaining monotonicity in the system size. The
critical line, as well as scaling of the magnetization and the
susceptibility, are also investigated obtaining full agreement
among theory and simulations.
\newline
Future works should extend these techniques to several lateral
models as the bipartite diluted mean field Ising models, while the
need of stronger techniques to go well beyond the critical line is
also to be satisfied as well as their practical application to
social science or biological networks. We plan to follow these
research lines in the future.

\section*{Acknowledgment}
The authors are pleased to thank Francesco Guerra, Pierluigi
Contucci and Peter Sollich for useful suggestions.
\newline
EA thanks the Italian Foundation ``Angelo della Riccia'' for
financial support.
\newline
AB acknowledges partial support by the CULTAPTATION project
(European Commission contract FP$6-2004$-NEST-PATH-$043434$) and
partial support by the MIUR within the Smart-Life Project
(Ministry Decree $13/03/2007$ n.$368$).

\end{document}